\documentclass[12pt]{article}

\usepackage{latexsym}
\usepackage{graphics}
\usepackage{amsfonts}
\usepackage{undertilde}

\newcommand{\be}{\begin{equation}}
\newcommand{\ee}{\end{equation}}
\newcommand{\ben}{\begin{eqnarray}}
\newcommand{\een}{\end{eqnarray}}
\newcommand{\bea}{\begin{eqnarray}}
\newcommand{\eea}{\end{eqnarray}}

\begin{document}

 \setlength{\baselineskip}{19pt}
\title{
\normalsize
\mbox{ }\hspace{\fill}
\begin{minipage}{7cm}
{\tt }{\hfill}
\end{minipage}\\[5ex]
{\large\bf Lagrangian formulation \\of the Palatini action
 \\[1ex]}}
\author{SangChul Yoon\footnote{scyoon@kunsan.ac.kr}
\\
\\
{\it Department of Physics, Kunsan National University},\\
{\it Kunsan 54150, Korea}\\
}

\maketitle

\thispagestyle{empty}

\begin{abstract}
We work on the Lagrangian formulation of the Palatini action. We
find that we must assume the metric compatibility condition for the
Palatini action to describe General Relativity, which condition
should hold in  quantization. We find that we must also assume one
of the torsion zero condition or the tetrad compatibility condition.
Our results will hold for any action in terms of the tetrad and the
internal connection which describes General Relativity.
\end{abstract}

Loop Quantum Gravity is a quantization program of General Relativity
based on Gauge Theory  \cite{Abers and Lee}. Although it has been
studied more than 30 years, yet it is still obscure what should be
assumed beforehand and what are derived afterward from the
Euler-Lagrange equations in the beginning Lagrangian formulation of
this program. In this paper, we clear this up once and for all. This
makes the Hamiltonian formulation richer than previously known.

The Palatini action is \bea S_p(e,w) \equiv \frac{1}{2}\int_M
\sqrt{-g} e^a_I e^b_J F_{ab}^{ \mbox{ } \mbox{ } IJ} \eea where
$e^a_I$ is a tetrad and $F_{ab}^{ \mbox{ } \mbox{ } IJ}$ is a
curvature tensor of an internal connection $w_{a}^{ \mbox{ } IJ}$:
\bea F_{abI}^{ \mbox{ } \mbox{ } \mbox{ }
J}=2\partial_{[a}w_{b]I}^{\mbox{ } \mbox{ }J} +[w_a, w_b]_I^{\mbox{
} J}.  \eea We can obtain a torsion tensor from $e^a_I$ and $w_{a}^{
\mbox{ } IJ}$: \bea T_{ab}^{\mbox{ }\mbox{ }I}=
2\partial_{[a}e_{b]}^I +w_{a \mbox{ } J}^{\mbox{ } I} e_b^J-w_{b
\mbox{ } J}^{\mbox{ } I} e_a^J.\eea

In the Palatini action, $e^a_I$ and $w_{a}^{ \mbox{ } IJ}$ are the
basic independent variables.  The Palatini action becomes the
Einstein-Hilbert action when the tetrad compatibility condition,
$D_a e^b_I=0$, holds where $D_a$ is a generalized derivative
operator defined on a generalized tensor field $H_{aI}$ by \bea D_a
H_{bI}\equiv
\partial_a H_{bI}+A_{ab}^{\mbox{ }\mbox{ }c} H_{cI} + w_{aI}^{\mbox{
}\mbox{ } J} H_{bJ}\eea where $A_{ab}^{\mbox{ }\mbox{ }c}$ is a
spacetime connection. In Riemannian geometry, $D_a e^b_I$ is always
zero and we can have the Riemann curvature tensor and the torsion
tensor from either $A_{ab}^{\mbox{ }\mbox{ }c}$ or $w_{aI}^{\mbox{
}\mbox{ } J}$, which makes two actions equivalent. In this case,
$A_{ab}^{\mbox{ }\mbox{ }c}$ and $w_{aI}^{\mbox{ }\mbox{ } J}$ are
determined by the tetrad and the torsion. From $D_a e^b_I=0$, $D_a
g_{bc}=0$. With this \bea A_{ab}^{\mbox{ }\mbox{ }c} = \Gamma_{ab}^{
\mbox{ } \mbox{ } c} + \frac{1}{2} \{ -T^{\mbox{ }c}_{a \mbox{ }
\mbox{ }b}- T^{\mbox{ }c}_{b \mbox{ }\mbox{ }a} +T_{ab}^{ \mbox{ }
\mbox{ } c} \} \eea where $ \Gamma_{ab}^{ \mbox{ } \mbox{ } c}$ is
the Christoffel symbols: \bea \Gamma_{ab}^{ \mbox{ } \mbox{ } c} =
-\frac{1}{2} g^{cd}\{
\partial_b g_{ad}+
\partial_a g_{bd} -\partial_d g_{ab} \}, \eea and  $T_{ab}^{ \mbox{ }
\mbox{ } c}$ is the spacetime torsion tensor defined by
non-commutativity of the derivative operator on a scalar field. With
$D_a e^b_I=0$ and (5), $w_{aI}^{\mbox{ }\mbox{ } J}$ are determined
by the tetrad and the torsion, which can be also obtained by solving
(3). If $D_a e^b_I \neq 0$, the Palatini action does not become the
Einstein-Hilbert action. The curvature and the torsion from (2) and
(3) are not equal to those obtained from $A_{ab}^{\mbox{ }\mbox{
}c}$. In this case, (2) and (3) do not have the geometrical meanings
of Riemannian geometry.

The variational calculations of the Palatini action are incorrect in
many literatures. There seems to have been confusion about what is
the correct statement on the relation between the metric
compatibility condition, the torsion zero condition and the tetrad
compatibility condition.  The correct statement is this: If
$D_ag_{bc}=0$ and $T_{ab}^{\mbox{ }\mbox{ }c}=0$, then
 $T_{ab}^{\mbox{ }\mbox{ }I}=0$ if and only if $D_a e^b_I=0$. Note that Stokes's theorem holds for
 a torsion-free derivative operator on a orientable manifold and
 Gauss's theorem holds  when the metric compatibility condition is satisfied once
a volume element is chosen by a metric. Because great care must
 be taken without $D_ag_{bc}=0$ or the torsion
 zero condition, let's work on a simple model first: \bea S \equiv \int_M \sqrt{-g} P^a Q^b
D_a R_b. \eea If $D_ag_{bc}=0$ and $T_{ab}^{\mbox{ }\mbox{ }c}=0$,
\bea
\partial_a(\sqrt{-g}P^a)=\sqrt{-g} D_a P^a \eea  where we used the
formula: \bea \partial_a \sqrt{-g}=\frac{1}{2}\sqrt{-g}g^{bc}
\partial_a g_{bc} =-\sqrt{-g}(A_{ba}^{\mbox{ }\mbox{ }b}-T_{ba}^{\mbox{ }\mbox{
}b}) .\eea Generally without assuming $D_ag_{bc}=0$, \bea
\partial_a(\sqrt{-g}P^a) = \sqrt{-g} D_a P^a+( \partial_a\sqrt{-g} +\sqrt{-g}
 A_{ba}^{\mbox{ }\mbox{ }b})P^a. \eea
 If $D_ag_{bc}=0$ but $T_{ab}^{\mbox{ }\mbox{ }c} \neq 0$,
\bea \partial_a(\sqrt{-g}P^a)=\sqrt{-g} D_a P^a+ \sqrt{-g}
T_{ba}^{\mbox{ }\mbox{ }b}P^a. \eea

Let's see what we have when we vary $R_a$. From $\delta S=0$, we
have   \bea \sqrt{-g} D_a (P^aQ^b) +( \partial_a\sqrt{-g} +\sqrt{-g}
A_{ca}^{\mbox{ }\mbox{ }c})P^aQ^b =0\eea where we used $\delta
R_a=0$ on the boundary. Note that the second term does not disappear
as in (10). If $D_ag_{bc}=0$, we have \bea  D_a (P^aQ^b)+
T_{ca}^{\mbox{ }\mbox{ }c}P^aQ^b=0. \eea We can see that a solution
$D_a (P^aQ^b)=0$  is obtained only when $D_ag_{bc}=0$ and
$T_{ab}^{\mbox{ }\mbox{ }c}=0$.

Let's work on the Palatini action (1) with the variational method.
To see what we have when we vary $e_I^a$, it is easier when we use
 \bea \tilde{\eta}^{abcd}
\epsilon_{IJKL}e^K_c e^L_d=-4\sqrt{-g}e^{[a}_I e^{b]}_J\eea where
$\tilde{\eta}^{abcd}$ is the Levi-Civita tensor density of weight 1
and $\epsilon_{IJKL}$ is the volume element of the Minkowski metric
$\eta_{IJ}$. Both of them are -1 or 0 or 1 depending on their
indices, so they are independent of $e_I^a$. With this, varying
$e_I^a$ gives \bea \tilde{\eta}^{abcd} \epsilon_{IJKL} e^J_b
F_{cd}^{ \mbox{ } \mbox{ } KL}=0. \eea For $w_{a}^{ \mbox{ } IJ}$,
we need the following formula: \bea \delta F_{ab}^{ \mbox{ } \mbox{
} IJ}=2D_{[a}\delta w_{b]}^{\mbox{ } IJ}-T_{ab}^{\mbox{ }\mbox{ }c}
\delta w_{c}^{\mbox{ }IJ}. \eea We can see immediately that the
variational calculations of the Palatini action with respect to
$w_{a}^{ \mbox{ } IJ}$ are very similar to those of our simple
action (7). If we assume $D_ag_{bc}=0$ and $T_{ab}^{\mbox{ }\mbox{
}c}=0$, varying $w_{a}^{ \mbox{ } IJ}$ gives us  \bea D_a( e^{[a}_I
e^{b]}_J)=0. \eea To figure out what (17) gives, we express $D_a$ in
terms of the unique, torsion-free generalized derivative operator
$\nabla_a$ compatible with $e^I_a$, and $C_{aI}^{\mbox{ }\mbox{ }
J}$ defined by \bea D_aH_I=\nabla_aH_I+C_{aI}^{\mbox{ }\mbox{ }
J}H_J.\eea If we express (17) with (18), it is not difficult to show
that $C_{aI}^{\mbox{ }\mbox{ } J}=0$. There are 24 homogeneous
linear equations of 24 variables $C_{aI}^{\mbox{ }\mbox{ } J}$, so
$C_{aI}^{\mbox{ }\mbox{ } J}=0$. Since $D_a=\nabla_a$, $F_{ab}^{
\mbox{ } \mbox{ } IJ}$ becomes the Riemann curvature tensor and we
have the 3+1 vacuum Einstein's equations from (15). If we do not
assume $D_ag_{bc} = 0$ but only assume $T_{ab}^{\mbox{ }\mbox{
}c}=0$, we have another term as in (10): \bea \sqrt{-g} D_a (
e^{[a}_I e^{b]}_J) +(\partial_a\sqrt{-g} +\sqrt{-g} A_{ca}^{\mbox{
}\mbox{ }c})( e^{[a}_I e^{b]}_J) =0.\eea If we express (19) with
(18), there are 24 inhomogeneous linear equations of 24 variables
$C_{aI}^{\mbox{ }\mbox{ } J}$, so $C_{aI}^{\mbox{ }\mbox{ } J} \neq
0$. In this case,  $D_a e^b_I$ is not zero. The Palatini action does
not become the Einstein-Hilbert action and we do not have the
Einstein's equations. Finally if we assume $D_ag_{bc}=0$ but do not
assume $T_{ab}^{\mbox{ }\mbox{ }c}=0$, varying $w_{a}^{ \mbox{ }
IJ}$ gives us \bea 2D_a( e^{[a}_I e^{b]}_J) +
e^a_Ie^c_JT_{ac}^{\mbox{ }\mbox{ }b}+e^a_{[I}e^b_{J]}T_{ca}^{\mbox{
}\mbox{ }c}=0.\eea We do not have $D_ae^b_I=0$ as the above and
therefore we do not have the Einstein's equations. If we assume
$D_ae^b_I=0$, we multiply $e^J_b$ to both sides and obtain
$T_{ac}^{\mbox{ }\mbox{ }c}=0$. Thus we have $T_{ab}^{\mbox{ }\mbox{
}c}=0$. In this case we have the Einstein's equations.

Since $D_ae^b_I=0$ means $D_ag_{bc}=0$,  we can see that we must
assume $D_ag_{bc}=0$ to have the Einstein's equations from the
Palatini action. Because this condition is assumed from the
beginning, it must be preserved in quantization. We also need to
assume either $T_{ab}^{\mbox{ }\mbox{ }c}=0$ or $D_ae^b_I=0$ to have
the Einstein's equations, which should be also preserved in
quantization. The conditions  $D_ag_{bc}=0$ and $T_{ab}^{\mbox{
}\mbox{ }c}=0$ are what Einstein assumed  when he constructed
General Relativity \cite{Ohanian and Ruffini}. With these two
conditions, geodesic is a extremal length between two spacetime
points, which is related to the Principle of Equivalence. On the
other hand, assuming $D_ae^b_I=0$ is based on Riemannian geometry.

Considering that  $T_{ab}^{\mbox{ }\mbox{ }c}$  and $w_{a}^{ \mbox{
} IJ}$ have the same number of components,  our results will hold
for any action in terms of $e^a_I$ and $w_{a}^{ \mbox{ } IJ}$ which
describes General Relativity. If we include the cosmological
constant and the standard model fields to the Palatini action, we
expect that the results are the same. Because two approaches are
equivalent classically, it is unlikely that only one approach
describes General Relativity. Our results will impose restrictions
on the standard model action in curved spacetime. Finally our
results make a clear distinction between the tetrad description and
the metric description of General relativity, which will be shown in
the Hamiltonian formulation. Although two descriptions are
equivalent classically, quantum theories are different. Two
approaches we found in the Lagrangian formulation will directly go
through the Hamiltonian formulation, giving a hint to solve 2nd
class constraints. This interplay between two formulations might
guide us to complete this program. We will work on these and
which approach leads to quantum gravity in future.\\

We thank Sang Pyo Kim for useful discussions and financial support.
This work was supported by Basic Science Research Program through
the National Research Foundation of Korea (NRF) funded by the
Ministry of Education (NRF-2015R1D1A1A01060626).\\

\end{document}